\documentclass[runningheads]{svmult}

\usepackage{makeidx}   
\usepackage{graphicx}  
\usepackage{subeqnar}  
\usepackage{multicol}  
\usepackage{physprbb}  
\makeindex             

\begin{document}
\title*{{\scriptsize \em Poster presented at the Hydrogen II workshop,
Castiglione della Pescaia, June 1 --3, 2000} \protect\newline ~~~
\protect\newline Hyperfine Structure Measurements of Antiprotonic
Helium and Antihydrogen}
%
%
\toctitle{Hyperfine Structure Measurements of Antiprotonic Helium
\protect\newline and Antihydrogen }
%
%
\titlerunning{Hyperfine Structure Measurements of $\overline{\mathrm p}$He$^+$ and $\overline{\mathrm H}$}
%
\author{Eberhard Widmann\inst{1}
\and John Eades\inst{2} \and Ryugo S. Hayano\inst{1}
\and Masaki Hori\inst{2}
\and Dezso Horvath\inst{3}
\and Takashi Ishikawa\inst{1}
\and Bertalan Juhasz\inst{4}
\and Jun Sakaguchi\inst{1}
\and Hiroyuki A. Torii\inst{5}
\and Hidetoshi Yamaguchi\inst{1}
\and Toshimitsu Yamazaki\inst{6}
}
\authorrunning{Eberhard Widmann et al.}
%
%
\institute{Department of Physics, University of Tokyo, Japan
\and CERN, Geneva, Switzerland
\and KFKI Research Institute for Particle and Nuclear Physics,
     Budapest, Hungary
\and University of Debrecen, Hungary
\and Institute of Physics, University of Tokyo, Japan
\and RI Beam Science Laboratory, RIKEN, Saitama, Japan
}

\maketitle              
\label{c_widm}
\setcounter{page}{508}

\begin{abstract}
This paper describes measurements of the hyperfine structure of
two antiprotonic atoms that are planned at the Antiproton
Decelerator (AD) at CERN. The first part deals with antiprotonic
helium, a three-body system of $\alpha$-particle, antiproton and
electron that was previously studied at LEAR. A measurement will
test existing three-body calculations and may -- through
comparison with these theories -- determine the magnetic moment
$\mu_{\overline{\mathrm p}}$\ of the antiproton more precisely
than currently available, thus providing a test of CPT invariance.
The second system, antihydrogen, consisting of an antiproton and a
positron, is planned to be produced at thermal energies at the AD.
A measurement of the ground-state hyperfine splitting
$\nu_{\mathrm{HF}}({\overline{\mathrm{H}}})$, which for hydrogen
is one of the most accurately measured physical quantities, will
directly yield a precise value for $\mu_{\overline{\mathrm p}}$,
and also compare the internal structure of proton and antiproton
through the contribution of the magnetic size of the
$\overline{\mathrm p}$\ to
$\nu_{\mathrm{HF}}({\overline{\mathrm{H}}})$.
\end{abstract}

\section{Introduction}
The upcoming Antiproton Decelerator (AD) \cite{46AD} at CERN
allows the formation and precision spectroscopy of antiprotonic
atoms. Among the three approved experiments, the ASACUSA
collaboration \cite{46ASACUSA} will as part of its program
continue experiments with antiprotonic helium that were previously
performed by the PS205 collaboration \cite{46PS205} at the now
closed Low Energy Antiproton Ring (LEAR) of CERN. Antiprotonic
helium consisting of an alpha particle, an antiproton, and an
electron (He$^{++}-\overline{p}-e^- \equiv$ $\overline{\mathrm
p}$He$^+$), was found to have lifetimes in the microsecond range,
thus enabling its examination with spectroscopy techniques. This
unusual 3-body system has both the properties of an  atom and --
due to the large mass of he $\overline{\mathrm p}$\ -- a molecule
and is therefore often called
 ``atomcule''. An overview on measurements on antiprotonic
helium is given in the talk by T.~Yamazaki \cite{46Yamazaki:here}.
The laser spectroscopy experiments of PS205 have proved that the
antiproton  occupies highly excited metastable states with
principal and angular quantum numbers $(n,L)$ = 30$\ldots$39 (cf.
Fig.~\ref{46fig:leveldiag}). A major experiment at the AD will be
the measurement of level splittings  caused by the magnetic
interaction of its constituents. Due to the large angular momentum
of the antiproton, the dominant splitting comes from the
interaction of the antiproton angular momentum and the electron
spin. Since it is caused by the interaction of different
particles, it is called a hyperfine structure (HFS). Its magnitude
is about 10$\ldots$15 GHz. The antiproton spin leads to a further,
by two orders of magnitude smaller splitting (called super
hyperfine structure, SHFS). We describe an already installed
two-laser microwave triple resonance experiment to determine this
unique level splitting accurately. A measurement of the HF
splitting will constitute a test of existing three-body
calculations and, through comparison with these calculations, has
the potential to become a CPT test by extracting a value of the
antiproton magnetic moment with possibly higher accuracy than it
is currently known.

The formation and spectroscopy of antihydrogen, the simplest form
of neutral antimatter consisting of an antiproton and a positron,
is one of the central topics at the AD. Complementary to the 1s-2s
laser spectroscopy pursued by the two other experiments at the AD,
ATHENA \cite{46ATHENA} and ATRAP \cite{46ATRAP}, the ASACUSA
collaboration is developing a measurement of the antihydrogen
ground-state hyperfine structure. This quantity is of great
interest for CPT studies in the hadronic sector, since this value
for hydrogen is one of the most accurately measured physical
quantities, but the theoretical precision is limited by the much
less accurately known electric and magnetic form factors of the
proton. By measuring the HFS of antihydrogen, the value of the
magnetic moment of the antiproton and its form factor, i.e. its
spatial distribution, can be compared to the ones of the proton.
Preliminary studies of a possible experimental layout using an
atomic beam method as employed in the early stages of the hydrogen
HFS measurements are presented.

\section{Hyperfine Structure of $\overline{\mathrm p}$He$^+$ }

\begin{figure}[t]
\begin{center}
\includegraphics[width=.6\textwidth]{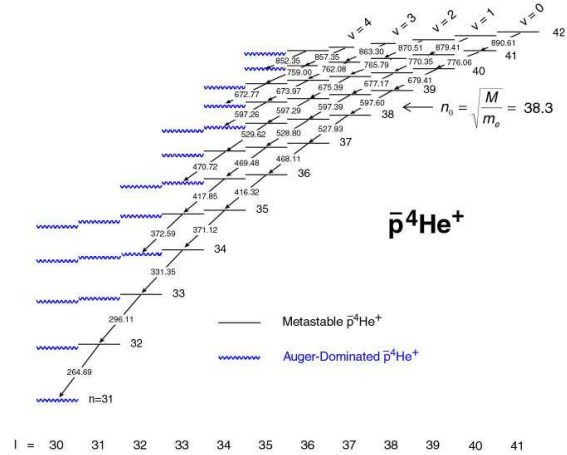}
\end{center}
\caption[]{Energy levels of the antiproton in $\overline{\mathrm
p}$He$^+$. The $\overline{\mathrm p}$\ is captured by replacing
one of the 1s electrons, which corresponds for the
$\overline{\mathrm p}$\ to a state with principal quantum number
$n_0 \sim \sqrt{M^*/m}$, where $M^*$ is the reduced mass of the
atomcule, and $m$ the electron mass. About 3\% of antiprotons are
captured in metastable states (black lines) at high angular
momenta $L\sim n-1$, for which deexcitation by Auger transitions
is much slower than radiative transitions. The lifetimes of these
states is in the order of $\mu$s. The antiprotons follow
predominantly cascades with constant vibration quantum number
$v=n-L-1$ (black arrows) until they reach an auger-dominated
short-lived state. The atomcule then ionizes within $<10$ ns and
the $\overline{\mathrm p}$He$^{++}$ \ is immediately destroyed in
the surrounding helium medium. The overall average lifetime of
atomcules is about $3-4$ $\mu$s.} \label{46fig:leveldiag}
\end{figure}

\begin{figure}[t]
\begin{center}
\includegraphics[width=.6\textwidth]{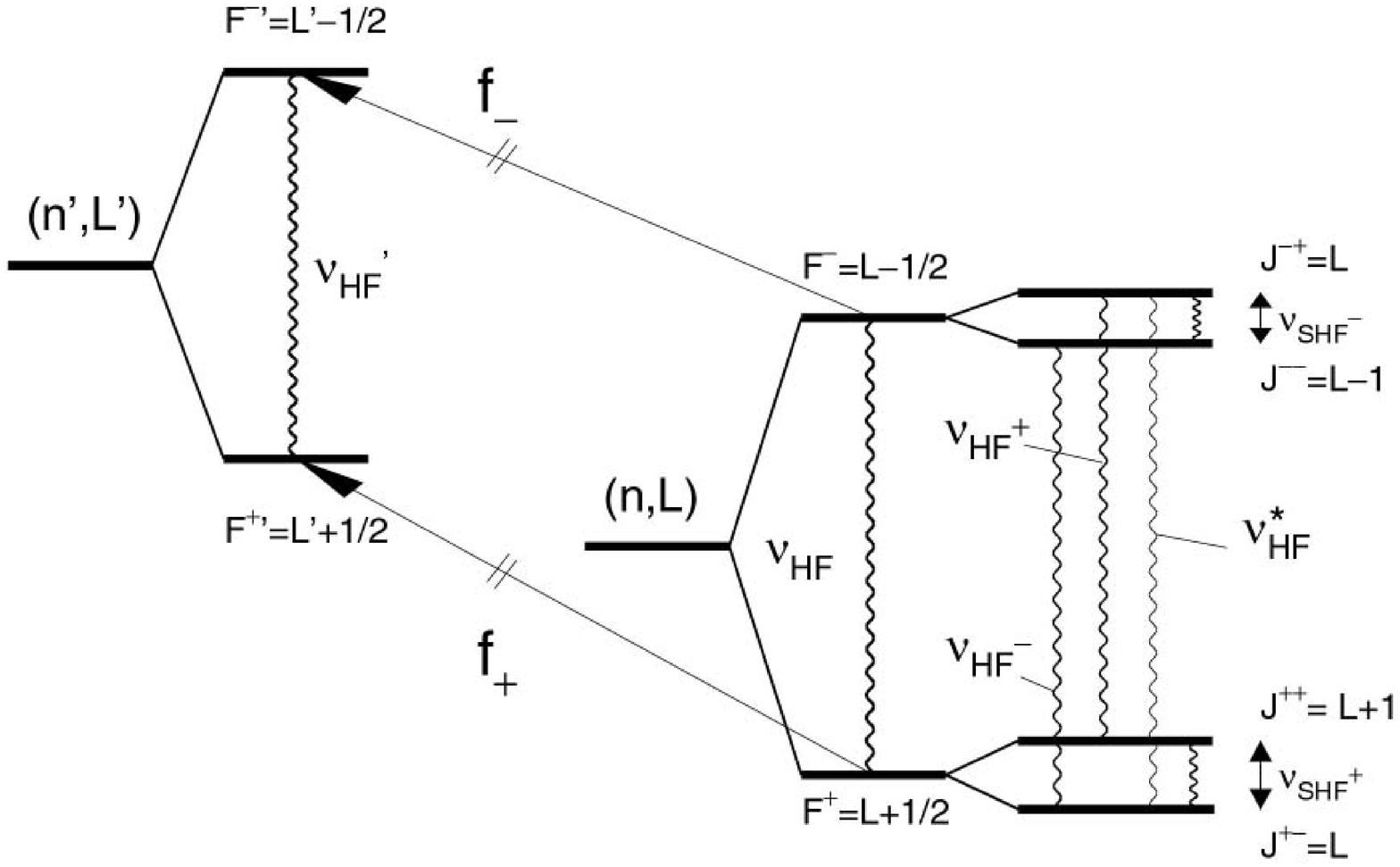}
  \caption{\label{46fig:hfs-leveldiag}Laser ($f^\pm$),
           microwave ($\nu_{\mathrm{HF}}^\pm$), and RF
           ($\nu_{\mathrm{SHF}}^\pm$)
           transitions in $\overline{\mathrm p}$He$^+$.
           $\nu_{\mathrm{HF}}^*$ is a
           transition between the SHF states of same total
           angular momentum $J=L$ that is suppressed by a factor
           $\sim 1/L$ compared to the allowed $\Delta L=1$ transitions
           $\nu_{\mathrm{HF}}^\pm$. The right-hand side
           shows the splitting of the parent state of a laser
           transition into a quadruplet, while on the left-hand
           side only the dominant doublet splitting for the
           daughter state is shown}
\end{center}
\end{figure}

Fig.~\ref{46fig:leveldiag} shows the level diagram of antiprotonic
helium which was experimentally established by observing several
laser-induced transitions of the antiproton (see talk by
T.~Yamazaki \cite{46Yamazaki:here}). Each level in
Fig.~\ref{46fig:leveldiag} is split due to the presence of three
angular momenta: the orbital angular momentum $L$ (mainly carried
by the $\overline{\mathrm p}$), and the spins of the electron
$S_e$ and the antiproton $S_p$. These momenta couple according to
the following scheme:
\begin{eqnarray}
\vec{F} &=& \vec{L} + \vec{S_e} \\ \vec{j} &=& \vec{L} + \vec{S_p}
\\ \vec{J} &=& \vec{F} + \vec{S_p} = \vec{j} + \vec{S_e} = \vec{L}
+ \vec{S_p} + \vec{S_e}
\end{eqnarray}
Due to the large orbital angular momentum of the
$\overline{\mathrm p}$\ ($L\sim 30\ldots35$) occupying metastable
states, the dominant splitting is caused by the interaction of the
spin-averaged $\overline{\mathrm p}$\ magnetic moment and the
electron spin, giving rise to a doublet with $F^+=L+1/2$ and
$F^-=L-1/2$ and the associated energy splitting $h\nu_\mathrm{HF}$
(cf. Fig.~\ref{46fig:hfs-leveldiag}). This splitting is called the
{\em Hyperfine (HF) Structure}. The $\overline{\mathrm p}$\ spin
causes an additional, smaller splitting for each of the HF states,
which is called here the {\em Super Hyperfine (SHF) Structure}.

The HF and SHF structure has been calculated by Bakalov and
Korobov \cite{46Bakalov1996a,46Bakalov-Korobov1998} using the best
three-body wavefunctions of Korobov \cite{46Korobov1997}, and
recently by Yamanaka et al. \cite{46Yamanaka:2000} using
wavefunctions calculated by Kino et al. by the coupled
re\-arrange\-ment-channel method \cite{46Kino:1999}. They present
the HF and SHF energies in terms of the angular momentum operators
as
\begin{eqnarray}
\delta E &=& E_1 (\vec{L} \cdot \vec{S_e}) + E_2 (\vec{L} \cdot
\vec{S_p}) \\ \nonumber & & + E_3 (\vec{S_e} \cdot \vec{S_p}) +
E_4 \{2L(L+1) (\vec{S_e} \cdot \vec{S_p})
 - 6(\vec{L} \cdot \vec{S_e}) \cdot (\vec{L} \cdot \vec{S_p})\}
\end{eqnarray}
\noindent The first term gives the dominant HF splitting.

The lower and the upper states of a HF doublet have $F^+$ and
$F^-$, respectively. The SHF structure is a combined effect of i)
(the second term) the one-body spin-orbit interaction (called
historically {\it Fine Structure}, but small in the present case,
because of the very large $(n,L)$), ii) (the third term) the
contact term of the $\vec{S_p}-\vec{S_e}$ interaction and iii)
(the fourth term) the tensor term of the $\vec{S_p}-\vec{S_e}$
interaction. According to the calculation, the contact and the
tensor terms almost cancel and the SHF splitting is nearly equal
to the one-body spin-orbit splitting as given by the second term.
Thus, its level order (the $j^- = L -1/2$ level is lower than the
$j^+ = L +1/2$ level) is therefore retained.

\subsection{Observation of a Line Splitting in a Laser Transition}

According to the previous chapter, one should observe several
lines in a single laser transition. Due to the large $L$, the
electric dipole transitions induced by the laser pulses are
subject to the selection rule $\Delta S_e= \Delta S_p=0$, which
results in a quadruplet structure of each transition line, where
the distance between the sub-lines is equal to the difference in
splittings of the parent and daughter states. But from theoretical
calculations \cite{46Bakalov-Korobov1998} it follows that the
splitting arising from the SHF structure is too small ($\approx
10\ldots50$ MHz) to be resolved in our experimental conditions
(the $\overline{\mathrm p}$\ of momentum 100 MeV/c (5.3 MeV
kinetic energy) are stopped in rather dense helium gas of
temperature $\sim 6$ K and pressure $\sim 250-600$ mbar) where the
Doppler broadening amounts to $\sim 400$ MHz. The splitting due to
the HF coupling, however, is in the order of $1.6\ldots1.9$ GHz
for so-called ``unfavoured'' transitions of type $\Delta v=2$
(transitions between different cascades, see
Fig.~\ref{46fig:leveldiag}) which is slightly larger than the
bandwidth of $\sim 1$ GHz of our laser system .

\begin{figure}[t]
         \begin{center}
         \includegraphics[width=.6\textwidth]{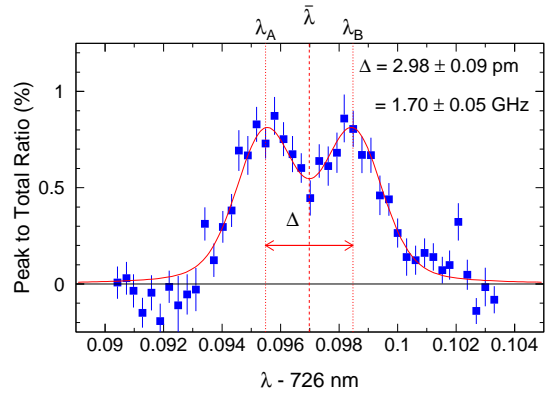}
          \end{center}
           \caption{Observed hyperfine splitting in the $(n,L)=
           (37,35)\rightarrow(38,34)$ transition of antiprotonic
           helium. Plotted here is the area under the laser-induced
           annihilation peak
          normalized to the total delayed annihilations vs. the
          laser wavelength}
    \label{46fig:HFS-obs}
\end{figure}

In the last beamtime at LEAR in 1996 we therefore scanned the
previously discovered $(n,L)=(37,35) \rightarrow (38,34)$
transition at $\lambda = 726.1$ nm by tuning our laser system to
the minimum achievable bandwidth of 1.2 GHz.
Fig.~\ref{46fig:HFS-obs} shows the result of the high resolution
scan \cite{46Widmann1997}: a doublet structure with a separation
of $\Delta \nu_\mathrm{HF}=1.70 \pm 0.05$ GHz, in accordance with
the theoretical prediction of Bakalov and Korobov of $1.77$ GHz
\cite{46Bakalov-Korobov1998}.

\subsection{Planned Two-Laser Microwave Triple Resonance Experiment}

Due to Doppler broadening and the limited bandwidth of pulsed
laser systems, the achievable accuracy in measuring the HF
splitting in a laser transition is rather small. Moreover, only
the difference of the splittings of parent and daughter state can
be measured. A more promising way is to directly induce
transitions between the HF and SHF levels within a state (e.g. the
transitions labeled $\nu_\mathrm{HF}^+$ and $\nu_\mathrm{HF}^-$ in
Fig.~\ref{46fig:hfs-leveldiag}) by applying microwave radiation.
According to \cite{46Bakalov-Korobov1998,46Yamanaka:2000}, the HF
splitting for the (37,35) state amounts to $\sim 12.91$ GHz.

\begin{figure}[b]
         \begin{center}
 \includegraphics[width=.6\textwidth]{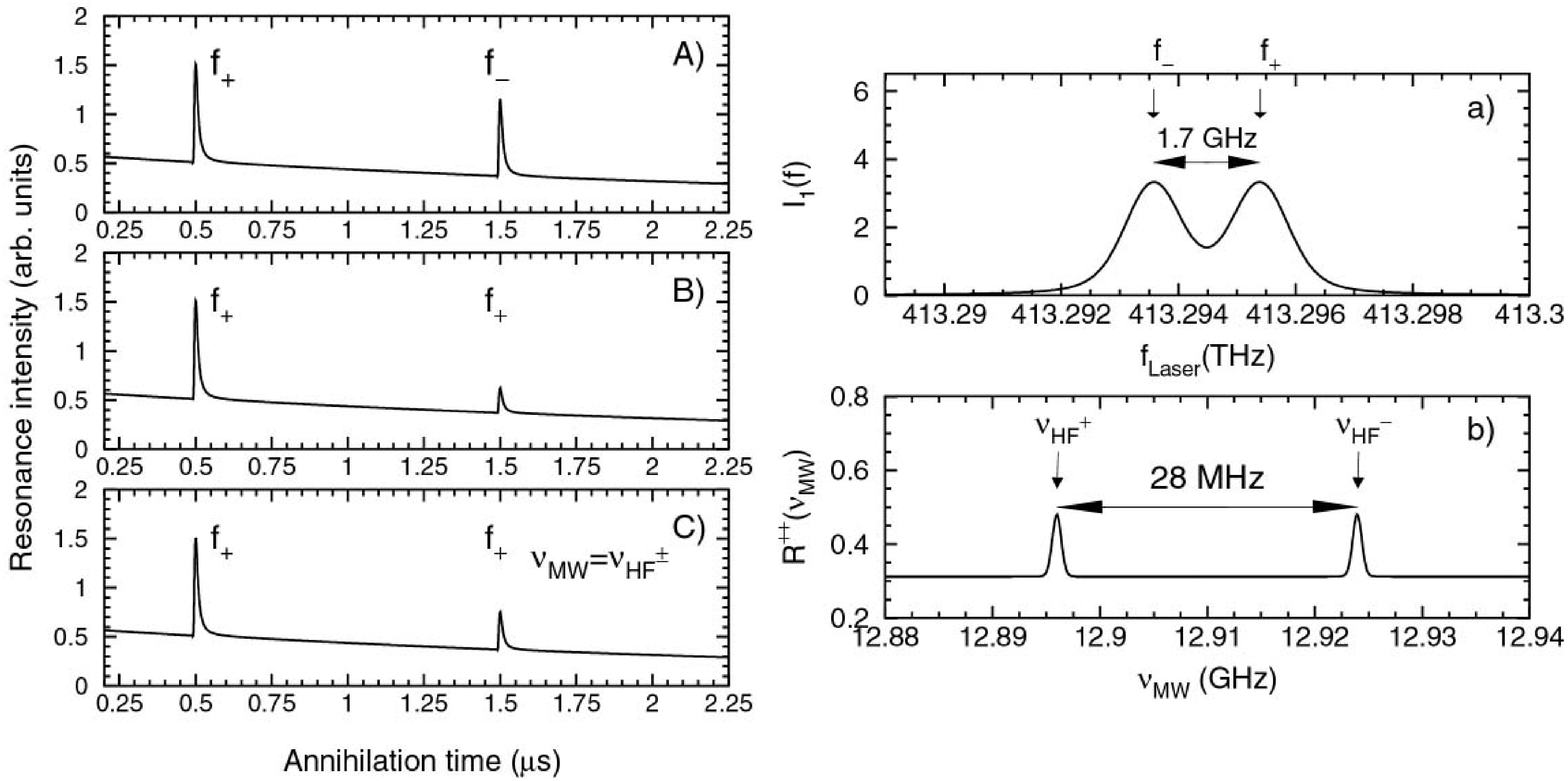}
          \end{center}
           \caption{Two-laser microwave triple resonance experiment
           explained at the example of the already observed
           $(n,L)=(37,35)\rightarrow(38,34)$ transition.
           Left:
           Simulated delayed annihilation time spectra of the
           laser/microwave triple resonance method. Right:
           Simulated laser and microwave resonance profiles}
    \label{46fig:3res-DATS}
\end{figure}

In order to detect a microwave induced transition between the
$F^+$ and $F^-$ states, first a population asymmetry has to be
induced. As described in Fig.~\ref{46fig:3res-DATS}, this can be
done by a laser pulse tuned to only the $f^+$ or $f^-$ transition.
Fig.~\ref{46fig:3res-DATS}a) shows a resonance profile of the
$(n,L)=(37,35)\rightarrow(38,34)$ laser transition assuming a
realistic laser bandwidth of $\sim 0.9$ GHz, with which the
doublet can be sufficiently separated. Nevertheless, a laser pulse
tuned to $f^-$ will still partly depopulate also the $F^+$ states.

Fig.~\ref{46fig:3res-DATS}A)-C) show time spectra when two
successive laser pulses are applied. In
Fig.~\ref{46fig:3res-DATS}A), the two pulses have different
frequencies, and therefore the pulse height is determined by the
population of the HF levels at times $t_1$ and $t_2$ when the
pulses are applied. In Fig.~\ref{46fig:3res-DATS}B) two pulses of
the {\em same} frequency are applied. In this case the laser peak
at $t_2$ is much smaller than the one at $t_1$, its height being
determined by the deexcitation efficiency of the first pulse, plus
feeding from upper states during the period between the two
pulses.

If between $t_1$ and $t_2$ a microwave pulse is applied on
resonance with one for the possible transitions between the $F^+$
and $F^-$ states (e.g. $\nu_\mathrm{HF}^+$, see
Fig.~\ref{46fig:hfs-leveldiag}), the population of these levels
can be equalized, and the second laser pulse at $t_2$ will detect
a larger population and thus the peak at $t_2$ will be larger. If
the ratio $R^{++}$ of the peak areas at $t_2$ and $t_1$ is plotted
against the microwave frequency, two resonances should be observed
for the two allowed transitions $\nu_\mathrm{HF}^+$ and
$\nu_\mathrm{HF}^-$, as shown in Fig.~\ref{46fig:3res-DATS}C),
with the center at 12.91 GHz and a splitting of
$\nu_\mathrm{SHF}^+ - \nu_\mathrm{SHF}^- = 28$ MHz as predicted by
\cite{46Bakalov-Korobov1998,46Yamanaka:2000}.

In this way, the HF splitting of $\overline{\mathrm p}$He$^+$
atomcules can be determined. The ultimate precision is limited by
the natural width of the metastable states ($\sim 0.2$ MHz) There
might, however, be distortions of the resonance line due to
influences of collisions with the surrounding medium that have so
far only been roughly estimated theoretically to yield only a
negligible shift and a broadening of $\sim 10$ MHz
\cite{46Korenman:1999}. In this case the line center could be
measured to $\sim 100$ kHz corresponding to $\sim 10$ ppm. The
accuracy of the theoretical predictions are $10^{-4}$ (100 ppm)
\cite{46Bakalov-Korobov1998} and 50 ppm \cite{46Yamanaka:2000},
and agree within 200 ppm ($2\times 10^{-4}$). The measurement will
therefore test the three-body calculations and QED corrections to
this accuracy.

The SHF frequencies provide information on the one-body spin-orbit
and spin-spin terms. If a doublet structure with a small splitting
is indeed observed in the two-laser microwave triple resonance
experiment, it confirms the cancellation of the scalar and tensor
spin-spin terms as predicted by theory. The observed difference of
$\nu_\mathrm{SHF}^+ -  \nu_\mathrm{SHF}^-$ is then rather
insensitive to the magnetic moment $\mu_{\overline{p}}$ of the
antiproton. An observation of the suppressed transition
$\nu_\mathrm{SHF}^*$ (see fig.~\ref{46fig:hfs-leveldiag}) or a
direct measurement of $\nu_\mathrm{SHF}^+$ or
$\nu_\mathrm{SHF}^-$, however, could reveal information on
$\mu_{\overline{p}}$ which is so far only known to $3\times
10^{-3}$ from X-ray measurements of antiprotonic Pb
\cite{46Kreissl}.

\section{Ground-State Hyperfine Structure of Antihydrogen}

\subsection{General Remarks}

The production and spectroscopy of antihydrogen
($\overline{\mathrm p}$--e$^+ \equiv$ $\overline{\mathrm H}$) is
one of the central topics at the Antiproton Decelerator (AD) of
CERN. The two other approved experiments, ATHENA \cite{46ATHENA}
and ATRAP \cite{46ATRAP} are dedicated to antihydrogen studies,
and plan to precisely measure the optical 1s--2s transition in
$\overline{\mathrm H}$\ using Doppler-free two-photon
spectroscopy. Tests of CPT symmetry performed by comparing
hydrogen and antihydrogen can yield unprecedented accuracy since
the ground state of antihydrogen has in principle an infinite
lifetime, if the $\overline{\mathrm H}$\ can be separated from
ordinary matter. Those experiments therefore plan to capture the
antihydrogen atoms in neutral atom trap as it has been done for
hydrogen by the group of D.~Kleppner
\cite{46Kleppner:here,46Cesar:1996}. Since the 2s state has a
natural linewidth of 1.3 Hz, they  hope to reach an ultimate
relative precision of $1 \times 10^{-18}$.

The 1s-2s transition energy is primarily due to the (electron)
Rydberg constant, where the antiproton mass contributes via the
reduced mass only of the order of $10^{-3}$. For the theoretical
calculations an uncertainty exists at the level of $5 \times
10^{-12}$ \cite{46Sapirstein} (finite size corrections) due to the
experimental error in the determination of the proton radius, even
if only the  more reliable Mainz value of $\sqrt{<r_p^2>} = 0.862
\pm 0.012$ fm~\cite{46chargeradius} is used (for a detailed
discussion of the proton radius  and its implications of precision
spectroscopy in hydrogen see~\cite{46Karshenboim:pradius}). It
should be noted that the recent determination of the hydrogen
ground-state Lamb shift from the 1s--2s transition energy
\cite{46Udem:97} favour an even slightly larger value of the
proton radius than stated above. In this sense the hydrogen and
antihydrogen 1s-2s energies yield primarily information on the
proton and antiproton charge distributions, respectively, once the
experimental accuracy exceeds the level of $5 \times 10^{-12}$.

The hyperfine structure of the ground state of the hydrogen atom
is also among the best known quantities in physics, which has had
a large impact on quantum physics at every stage of its
development, as described in a review by Ramsey \cite{46Ramsey}.
The first measurements were done 60 years \cite{46Rabi} ago with
Stern-Gerlach type inhomogeneous magnets where the hydrogen atoms
were deflected by the force of the magnetic field gradients onto
the magnetic moment of the electron.

With the advent of the magnetic resonance method the hyperfine
splitting of the hydrogen ground state was successfully determined
from microwave resonance transitions  by Nafe and Nelson
\cite{46Nafe} and later by Prodell and Kusch \cite{46Prodel}. The
precision attained by the first resonance experiment was already
impressively high: $\nu_{\mathrm{HF}}({\overline{\mathrm{H}}})$\
$= 1420.410 \pm 0.002$ MHz, which supersedes the uncertainty in
the theoretical prediction of the present day, as shown below. In
the early stages, room temperature hydrogen atoms were transported
through inhomogeneous magnetic field and the transit time in the
resonance cavity set an intrinsic limit on the resonance width. As
hydrogen atoms became confined, the precision increased
accordingly, and even a maser oscillation  was finally observed
\cite{46Goldenberg60}. The best value to the present
\cite{46Ramsey,46Essen:71,46Hellwig:70} is

\begin{equation}
\nu_\mathrm{HF}(\overline{\mathrm H}) = 1~420~405~751.7667 \pm
0.0009 ~{\rm Hz}.
\end{equation}

In the case of antihydrogen we can trace a similar historical
development, starting from a Stern-Gerlach type experiment and
proceeding to microwave resonance experiments. At each stage a
meaningful value of the antihydrogen hyperfine structure constant
can be obtained. The possible results and the achievable precision
are discussed in the next section. Section \ref{46sec:Hbarexp}
describes a possible experimental scenario to measure
$\nu_{\mathrm{HF}}({\overline{\mathrm{H}}})$.

\subsection {Hydrogen Hyperfine Structure and Related CPT
Invariant Quantities}

The hyperfine coupling frequency in the hydrogen ground state is
given to the leading term by the Fermi contact interaction,
yielding

\begin{equation} \label{46eq:fermi}
\nu_\mathrm{F} = \frac{16}{3} (\frac{M_p}{M_p+m_e})^3
\frac{m_e}{M_p} \frac{\mu_p}{\mu_N} \alpha^2 c Ry\\,
\end{equation}
which is  a direct product of the electron magnetic moment and the
anomalous proton magnetic moment (here $\hbar = c = 1$). Using the
known proton magnetic moment, this formula yields $\nu_\mathrm{F}
= 1418.83$ MHz, which is significantly different from the
experimental value and subsequently led to the discovery of the
anomalous electron $g$-factor.

Even after higher-order QED corrections \cite{46Sapirstein} still
a significant difference between theory and experiment remained,
as

\begin{equation}
\delta ({\rm QED}) = \frac{\nu ({\rm QED}) - \nu ({\rm Exp})}{\nu
({\rm Exp})} = 32.55 (10)  ~{\rm ppm}.
\end{equation}

This discrepancy  was largely accounted for by the
non-relativistic magnetic size correction (Zemach correction)
\cite{46Sapirstein}:

\begin{equation}
\Delta \nu ({\rm Zemach}) = \nu_{\mathrm F}
    \frac{2 Z \alpha m_{\mathrm e}}{\pi^2}
    \int{
    \frac{d^3p}{p^4}
    \left[ \frac{G_E(p^2)G_M(p^2)}{1+\kappa} -1
    \right]
    },
\end{equation}

where $\nu_{\mathrm F}$ is the Fermi contact term defined in
(\ref{46eq:fermi}), $G_E(p^2)$ and $G_M(p^2)$ are the electric and
magnetic form factor of the proton, and $\kappa$ its anomalous
magnetic moment. The Zemach corrections therefore contain both the
magnetic and charge distribution of the proton.

A detailed treatment of the Zemach corrections can be found in
\cite{46Karshenboim:Zemach}. Assuming the validity of the dipole
approximation, the two form factors can be correlated
\begin{equation}\label{46eq:formf}
  G_E(p^2) = \frac{G_M(p^2)}{1+\kappa}
   = \left(\frac{\Lambda^2}{\Lambda^2+p^2} \right)^2
\end{equation}

where the $\Lambda$ is related to the proton charge radius by $r_p
= \sqrt{12}/\Lambda$. Whether the dipole approximation is indeed a
good approximation, however, is not really clear. Integration by
separation of low and high-momentum regions with various
separation values, and the use of different values for $r_p$ gives
a value for the Zemach corrections of $\Delta \nu(\mathrm{Zemach})
= -41.07(75)$ ppm \cite{46Karshenboim:Zemach}. With this
correction, and some more recently calculated ones, the
theoretical value deviates from the experimental one by
\cite{46Karshenboim:Zemach}

\begin{equation}
\frac{\nu(\mathrm{exp}) -
     \nu(\mathrm {th})}
    {\nu(\mathrm {exp})} = 3.5 \pm 0.9 \; \mathrm{ppm}.
\end{equation}

A further structure effect, the proton polarizability, is only
estimated to be $< 4$ ppm \cite{46Karshenboim:Zemach}, of the same
order than the value above. The ``agreement'' between theory and
experiment is therefore only valid on a level of $\sim 4$ ppm.
Thus, we can say that the uncertainty in the hyperfine structure
reflects dominantly the electric and magnetic distribution of the
proton, which is related to the origin of the proton anomalous
moment, being a current topics of particle-nuclear physics.

The hyperfine structure of antihydrogen  gives unique information,
which is qualitatively different from those from the binding
energies of antihydrogen. As the hyperfine coupling constants of
hydrogen and deuterium provided surprisingly anomalous values of
the proton and deuteron magnetic moments in the history of
physics, the measurement of antihydrogen hyperfine structure will,
first of all, give a value of the antiproton magnetic moment
($\mu_{\overline{p}}$), which is poorly known to date (0.3 \%
relative accuracy) from the fine structure of a heavy antiprotonic
atoms \cite{46Kreissl}. Furthermore, a precise value of
$\nu_\mathrm{HF}(\overline{\mathrm H})$ will yield information on
the magnetic and charge radius of antiproton.

\subsection{Proposed Experimental Scenario to Measure the
Ground-state Hyperfine Structure of Antihydrogen}
\label{46sec:Hbarexp}

\begin{figure}[b]
\begin{center}
 \includegraphics[width=.6\textwidth]{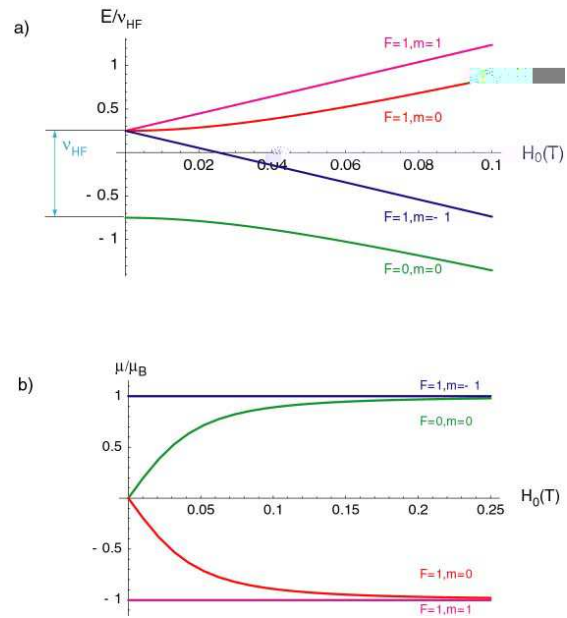}
\end{center}
\caption{\label{46fig:breit-rabi} a) Breit-Rabi diagram of the
hydrogen ground state in a magnetic field.  b) dependence of the
magnetic moment of he four states on the magnetic field }
\end{figure}

First it is important to remember the basics of the ground state
of hydrogen in a magnetic field. The two spin 1/2 particles proton
and electron (or antiproton and positron) can combine to states
according to $\vec{F}=\vec{S_\mathrm p}+\vec{S_\mathrm e}$ with
total spin $F$ = 0 or 1. The $F=1$ state has three possible
projections to a magnetic field axis described by the quantum
number $M_F=1, 0, -1$. At zero external field these three states
are degenerate, but at non-zero magnetic field their energy
evolves according to the well-known Breit-Rabi diagram
Fig.~\ref{46fig:breit-rabi}a). The two states  $(F,M_F)=(0,0)$ and
$(1,0)$ have no resulting magnetic moment at zero external field,
but develop one with increasing magnetic field due to the
decoupling of the two spins (Fig.~\ref{46fig:breit-rabi}b).

In the following the different steps of the experiment are
described.

\begin{figure}[b]
\begin{center}
 \includegraphics[width=.6\textwidth]{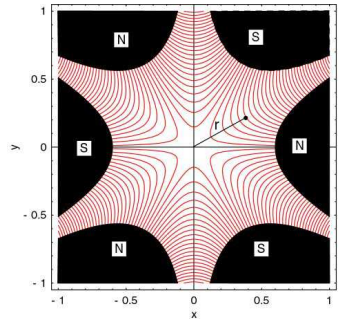}
\end{center}
\caption{\label{46fig:sextupole} Cross section of a sextupole
magnet with magnetic field lines (red). The direction of the atoms
is perpendicular to the cross section}
\end{figure}

\subsection*{Antihydrogen Formation}

The first step towards a measurement of
$\nu_{\mathrm{HF}}({\overline{\mathrm{H}}})$\ is the formation of
Antihydrogen. Here we assume the formation scheme and parameters
of the ATHENA \cite{46ATHENA-form}  experiment at CERN/AD.
\begin{itemize}
  \item $\overline{\mathrm H}$\ is formed form clouds of antiprotons and positrons
  trapped in Penning traps.
  \item the formation will be done by pushing antiprotons through
  a rotating positron plasma. The rotation is an unavoidable result of
  the $\vec{E} \times \vec{B}$ drift of the positrons in the magnetic field.
  The rotation frequency depends on the spatial density of the
  plasma \cite{46plasmarot}.
  \item the antiprotons will stop in the e$^+$-plasma and begin
  thermal diffusion until they form $\overline{\mathrm H}$.
  \item the $\overline{\mathrm H}$\ atoms will not be confined by the constant
  solenoid field and therefore leave the trap region with an
  energy distribution given by the temperature of the e$^+$-plasma
  and its rotation speed.
\end{itemize}

\subsection*{Transport and Spin Selection by Inhomogeneous
Magnetic Fields}

Since the $\overline{\mathrm H}$\ atoms will leave the solenoid
not as a collimated beam, it is straight forward to use sextupole
magnets to focus them as it was commonly done it atomic beam
experiments \cite{46atombeam}. The sextupole magnet will at the
same time act as a filter to select one of the two hyperfine
states. This can bee seen from writing the potential energy $V$ of
a magnetic moment of a spin-1/2 particle in a magnetic field and
the resulting force $K$:
\begin{eqnarray}\label{46eq:sext}
  V &=& - \vec{\mu} \cdot \vec{H(\vec{x})} \\ \nonumber
\vec{K} &=& - \nabla  V = \pm \mu \; \nabla H,
\end{eqnarray}
where the sign of $K$ depends on whether $\vec{\mu}$ and $\vec{H}$
are parallel or antiparallel. From Fig.~\ref{46fig:breit-rabi}b)
it is clear that the states $(F,M_F)=(1,1)$ and $(1,0)$ prefer
lower magnetic fields since this minimizes their energy
(``low-field seekers''), while the opposite is true for the other
two states.

\begin{figure}[b]
\begin{center}
 \includegraphics[width=.5\textwidth]{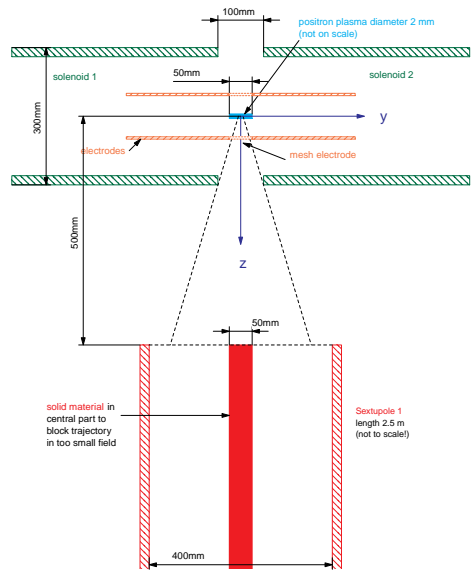}
\end{center}
\caption{\label{46fig:Hbar-magnet} Layout of the split solenoid
magnet for antihydrogen production and extraction}
\end{figure}

For a sextupole, the magnetic field is proportional to the square
of the radius, and the force therefore becomes proportional to $r$
(cf. Fig.~\ref{46fig:sextupole}):
\begin{equation}\label{46eq:force}
  H_{\mathrm {sextupole}} =\frac{C}{2}\, r^2 \rightarrow K_{\mathrm
  {sextupole}}= \pm C \mu r.
\end{equation}
For the (1,1) and (1,0) states this force points towards the
center of the sextupole and acts like the restoring force of a
harmonic oscillator for atoms leaving the center line.  This leads
to a harmonic oscillation in $r$, perpendicular to the propagation
direction. Atoms with the same velocity will therefore undergo
point-to-point focusing, where the focal length depends on the
velocity.

Fig.~\ref{46fig:Hbar-magnet} shows a realistic layout of a Penning
trap in a split solenoid magnet, where a sextupole magnet is
placed under 90 degrees to the solenoid magnetic field axis. The
splitting of the solenoid as well as a split electrode or one made
out of a mesh are necessary to let the $\overline{\mathrm H}$\
atoms pass without destroying them.

The further layout of the experiment consists of a first sextupole
S1 as a spin selector, a microwave cavity, and a second sextupole
S2 as a spin analyzer (cf. Fig.~\ref{46fig:Hbar-trajectories}).
Since the magnetic field in S1 is higher on the outside, the
``low-field seekers'' (1,1) and (1,0) will be focused, while the
``high-field seekers'' will move towards the magnet poles and
annihilate there. We further assume that the states with $M_F=0$
that have no permanent magnetic moment will loose their
orientation in the field-free region before S1, so that after S1
only atoms in the state (1,1) will remain.

\subsection*{Microwave Induced Spin-Flip Transition and Detection}

A microwave cavity placed between S1 and S2 can induce spin-flip
transitions $(F,M_F)=(1,1)\rightarrow(1,-1)$ if tuned to
$\nu_{\mathrm{HF}}({\overline{\mathrm{H}}})$. In order to produce
a positive signal, i.e. an {\em increase} in counting rate after
S2 under resonance condition, S2 will be rotated by 180 degrees
with respect to S1. Therefore, the $(1,-1)$ state where $M_F=-1$
is defined with respect to the magnetic field direction in S1 will
be a (1,1) state in S2, while the (1,1) state of S1 without spin
flip would correspond to a $(1,-1)$ state in S2. As a result, if
the microwave frequency is off resonance, no $\overline{\mathrm
H}$\ atoms will reach behind S2, while on resonance an increase in
the number of atoms should be detected after S2.

\begin{table}[b]
\begin{center}
\caption{\label{46tab:param} List of parameter used in the
simulation of $\overline{\mathrm H}$\ trajectories}
\begin{tabular}{lrl} \hline
  Source parameters & value & ~~~comment \\ \hline
  internal temperature of particle clouds & 10 K  \\
  rotation frequency of e$^+$ plasma & 100 kHz & ~~~for $10^8$ e$^+$/cm$^3$ \\
  diameter of e$^+$ plasma & 2 mm  & ~~~FWHM gaussian \\
  length of e$^+$ plasma&  5 cm \\ \hline
  Sextupole parameters\\ \hline
  max. field at pole & 1 T \\
  outer diameter sextupole & 40 cm\\
  inner diameter sextupole & 5 cm& ~~~blocked to reduce background\\
  & & ~~~of $\overline{\mathrm H}$\ atoms hitting directly\\
  & & ~~~the counter\\ \hline
  Microwave cavity\\ \hline
  typical dimensions  & 21 cm & ~~~wavelength for $\nu=1.4$ Ghz\\ \hline
\end{tabular}
\end{center}
\end{table}

The achievable  resolution for
$\nu_{\mathrm{HF}}({\overline{\mathrm{H}}})$ in this type of
experiment is determined by the flight-time of the atoms through
the cavity. Taking the average velocity of the atoms  of 650 m/s
(determined by the assumptions given in Table~\ref{46tab:param})
and a typical cavity length of 10 cm, a width of the resonance
line of $\sim 7$ kHz results corresponding to a fraction of $5
\times 10^{-6}$ relative to
$\nu_{\mathrm{HF}}({\overline{\mathrm{H}}})$ = 1.4 GHz. Using a
cavity of 50 cm length the width of the resonance line will be
reduced to 1 ppm. With enough statistics the center of the line
can be determined to about 1/100 yielding an ultimate precision of
$\sim 10^{-8}$.

\subsection*{Monte-Carlo Simulation of Proposed Experiment }

In order to estimate the efficiency of the setup as described
above, a Monte Carlo simulation was performed for the whole
experiment by numerically integrating the equation of motion
\begin{equation}\label{46eq:motion}
  \vec{K}= m \frac{\mathrm d^2 \vec{x}}{\mathrm{d}t}= -
  \mu_{\mathrm{eff}}(H(\vec{x})) \frac{\mathrm{d}
  H(\vec{x})}{\mathrm{d}\vec{x}},
\end{equation}
with $\vec{x}=(x,y,z)$ being the position of the atoms.

\begin{figure}[t]
\begin{center}
 \includegraphics[width=.7\textwidth]{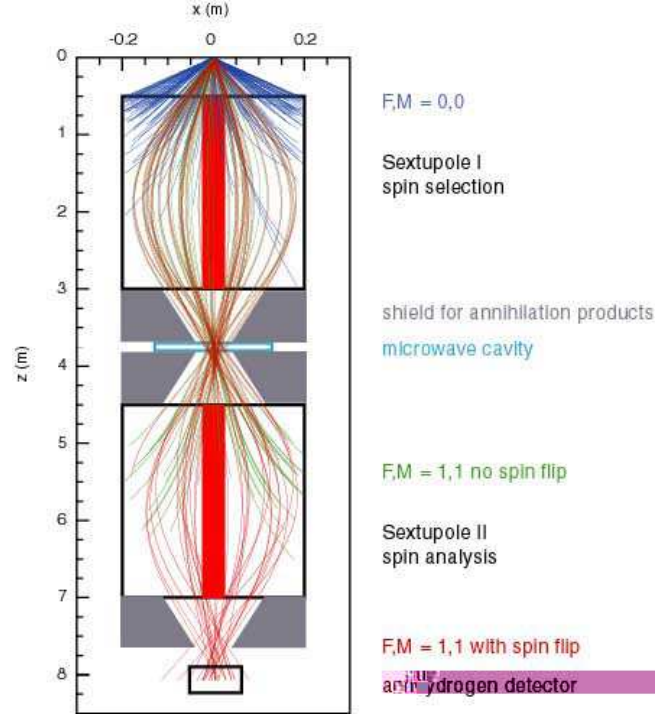}
\end{center}
\caption{\label{46fig:Hbar-trajectories} Trajectories of
antihydrogen atoms through two sextupole magnets as calculated by
a Monte-Carlo method. Please not the the X and Z axis do not have
he same scale }
\end{figure}

The magnetic field of the split solenoid was calculated by
numerically solving the Biot-Savart law for the geometry as
described in Fig.~\ref{46fig:Hbar-magnet} and a field of about 3~T
at the center. Inside the sextupole magnet the magnetic field was
calculated according to the analytical formula $H(r)\propto r^2$
with $r^2=x^2+y^2$, assuming a maximum field of $H(r_\mathrm{max})
= 1$ Tesla ($r_\mathrm{max}$ being the radius of the sextupole
magnets). The main assumptions are summarized in
Table~\ref{46tab:param}.

Typical trajectories calculated by this Monte Carlo method are
shown in Fig.~\ref{46fig:Hbar-trajectories}. The overall result is
that about $7 \times 10^{-5}$ of all antihydrogen atoms initially
formed in the trap region can be transported to the
$\overline{\mathrm H}$\ detector after S2. At expected formation
rates of about 200/s \cite{46ATHENA-form} this would result in a
count rate of 1 event per 2 minutes on resonance. This seems
rather small, but is feasible since the antihydrogen atoms can be
easily detected with unity efficiency from the annihilation of
their constituents.

\section{Conclusion and Summary}

The experiments with antiprotonic atoms discussed in this paper
have different main topics. The measurement of the hyperfine
structure of antiprotonic helium, which is already in progress at
the AD at CERN, will primarily test the accuracy of 3-body
calculations in the extreme situation of one particle having an
angular momentum quantum number of $\sim 35$. This makes it a very
difficult and challenging problem to few-body theory. If the
experimental accuracy could become high enough, we would be able
to perform a test of CPT theory for the magnetic moment of proton
and antiproton by comparing the experimental result to the
calculations that use the much better known value of the magnetic
moment of the proton. The equality of proton and antiproton
magnetic moment is experimentally only know to an accuracy of
0.3\%.

A measurement of the ground-state hyperfine structure of antihydrogen
would primarily be a CPT test in the hadronic sector, since the leading
term is directly proportional to the magnetic moment of the antiproton.
It is therefore a complementary measurement to the proposed 1s-2s
spectroscopy of antihydrogen, since the optical spectrum of
antihydrogen is dominated by the positron mass. The experimental value
for the hyperfine structure of hydrogen has for 30 years been one of
the most precisely measured value in physics (uncertainty $< 10^{-12}$)
and has only recently been surpassed by the 1s-2s two-photon laser
spectroscopy. On the other hand there exists an uncertainty in the
theoretical calculations on the level of $\sim 4$ ppm due to the finite
size of the proton, {\em i.e.} its electric and  magnetic structure and
polarizability. A determination of the hyperfine splitting of
antihydrogen with higher precision will therefore give insight into the
internal structure of the antiproton in comparison to the proton, which
is an actual topic in nuclear and high energy physics in relation to
the origin of the anomalous magnetic moment of the proton. The
experiment sketched in this paper seems feasible provided the
assumptions on the circumstances of antihydrogen production are
reasonable. More will be known about these when the first cold
antihydrogen atoms will be produced at the AD, hopefully this year or
in the year 2001. With enough statistics a resolution well below the
ppm level can be easily reached for the hyperfine splitting and
therefore the antiproton magnetic moment.

%
\label{c_widm_}

\begin{thebibliography}{8.}
\addcontentsline{toc}{section}{References}

\bibitem{46AD} AD homepage: \verb+http://www.cern.ch.PSdoc/acc/ad/+
\bibitem{46ASACUSA} T. Azuma et al: CERN/SPSC 97-19, CERN/SPSC
2000-04; ASACUSA home page \verb+http://www.cern.ch/ASACUSA/+
\bibitem{46PS205} PS205 home page:
\verb+http://www.cern.ch/LEAR_PS205/+
\bibitem{46Yamazaki:here} T. Yamazaki: {\em this edition}
\bibitem{46ATHENA} C. Amsler et al.: {\em this edition},
        pp. 449--468
\bibitem{46ATRAP} J. Walz et al.: {\em this edition},
        pp. 501--507
\bibitem{46Bakalov1996a} D.D. Bakalov  et al.:
        Phys. Lett. A {\bf 211}, 223 (1996)
\bibitem{46Bakalov-Korobov1998} D.D. Bakalov and V.I. Korobov:
        Phys. Rev. A {\bf 57}, 1662 (1998)
\bibitem{46Korobov1997} V.I. Korobov and D.D. Bakalov:
        Phys. Rev. Lett.  {\bf 79}, 3379 (1997)
\bibitem{46Yamanaka:2000} N. Yamanaka, Y. Kino, M. Kamimura, and
        H.~Kudo: Phys. Rev. A, in print
\bibitem{46Kino:1999} Y. Kino, M. Kamimura, and H. Kudo:
        Hyperfine Interaction {\bf 119}, 201 (1999)
\bibitem{46Widmann1997} E. Widmann at al.:
        Phys. Lett.  B {\bf 404}, 15 (1997)
\bibitem{46Korenman:1999} G. Korenman: private communication (1999)
\bibitem{46Kreissl}  A. Kreissl et al.:
        Z. Phys. C {\bf 37}, 557 (1988)
\bibitem{46Kleppner:here} D. Kleppner: {\em this edition},
        pp. 29--43
\bibitem{46Cesar:1996}
        C. Cesar {\it et al.}: Phys. Rev. Lett. {\bf77}, 255
        (1996)
\bibitem{46Sapirstein} J.R. Sapirstein and D. R. Yennie: `Theory
        of Hydrogenic Bound States'. In: {\em Quantum
        Electrodynamics}, ed. by T. Kinoshita (World Scientific,
        Singapore 1990) pp. 560--672
\bibitem{46chargeradius} G.G. Simon, Ch. Schmitt, F. Borokowski
        and V.H. Walther: Nucl. Phys. A {\bf 333}, 381 (1980)
\bibitem{46Karshenboim:pradius} S. G. Karshenboim:
        Can. J. Phys. {\bf 77}, 241 (1999)
\bibitem{46Udem:97} Th. Udem {\it et al.}:
        Phys. Rev. Lett. {\bf 79}, 2646 (1997)
\bibitem{46Ramsey} N. Ramsey: `Atomic Hydrogen Hyperfine Structure
        Experiments'. In: {\em Quantum Electrodynamics},
        ed. by T. Kinoshita (World Scientific, Singapore 1990)
        pp. 673--695
\bibitem{46Rabi} I.I. Rabi, J.M.B. Kellogg and J.R. Zacharias:
        Phys. Rev. {\bf 46}, 157 and 163 (1934);
        J.M.B. Kellogg, I.I. Rabi and J.R. Zacharias: Phys. Rev.
        {\bf 50}, 472 (1936)
\bibitem{46Nafe} J.E. Nafe and E.B. Nelson: Phys. Rev. {\bf 73}, 718 (1948)
\bibitem{46Prodel} A.G. Prodell and P. Kusch: Phys. Rev. {\bf 88}, 184 (1952)
\bibitem{46Goldenberg60} H.M. Goldenberg, D. Kleppner and N.F.  Ramsey: Phys.
        Rev. Lett. {\bf 8}, 361 (1960)
\bibitem{46Essen:71} L. Essen, R.W. Donaldson, M.J. Bangham and E.G.
        Hope: Nature {\bf 229}, 110 (1971)
\bibitem{46Hellwig:70} H. Hellwig et al.: Proc. IEEE Trans. IM-19,
        200 (1970)
\bibitem{46Karshenboim:Zemach} S. G. Karshenboim:  Phys. Lett A
        {\bf 225}, 97 (1997)
\bibitem{46H1s-2s} M. Niering {\it et al.}: Phys. Rev. Lett. {\bf
        84}, (2000)
\bibitem{46HLS} S.R. Lundeen and F.M. Pipkin: Phys. Rev. Lett. {\bf
        46}, 232 (1981)
\bibitem{46RPP} D.E. Groom {\it et al.}: Europ. Phys. J. C {\bf
        15}, 1 (2000)
\bibitem{46ATHENA-form} R. Landua, ATHENA spokesman: private
        communication (2000)
\bibitem{46atombeam} P. Kusch and V.W. Hughes: `Atomic and
        Molecular Beam Spectroscopy'. In: {\em Encyclopedia of Physics Vol.
        XXXVII/1}, ed. by S. Fl\"ugge (Springer, Berlin 1959) pp.
        1--172
\bibitem{46plasmarot}F. Anderegg, E.M. Hollmann, and C.F.
        Driscoll: Phys. Rev. Lett. {\bf 81}, 4875 (1998)
\end{thebibliography}
\end{document}